\documentclass[aps,showpacs,showkeys]{revtex4-1}

\newcommand{\sect}[1]{\setcounter{equation}{0}\section{#1}}

\newcommand{\subsect}[1]{\setcounter{equation}{0}\subsection{#1}}

\usepackage{graphicx}

\begin{document}
\title{The $\kappa$-statistics approach to epidemiology}

\author{Giorgio Kaniadakis $^{1,}$*
, Mauro M. Baldi $^2$
, Thomas S. Deisboeck $^3$
, Giulia Grisolia $^4$
, Dionissios T. Hristopulos $^5$
, Antonio M. Scarfone $^6$
, Amelia Sparavigna $^1$
, Tatsuaki Wada $^7$
 and Umberto Lucia $^4$
}

\vspace{5mm}

\affiliation{
$^1$ Dipartimento di Scienza Applicata e Tecnologia, Politecnico di Torino, Corso Duca degli Abruzzi 24, 10129 Torino, Italy\\
$^2$ Dipartimento di Informatica, Sistemistica e Comunicazione, Università di Milano-Bicocca, Viale Sarca, 336 - 20126, Milano, Italy\\
$^3$ Department of Radiology, Harvard-MIT Martinos Center for Biomedical Imaging, Massachusetts General Hospital and Harvard Medical School, Charlestown, MA 02129, USA\\
$^4$ Dipartimento di Energia ``Galileo Ferraris'', Politecnico di Torino, Corso Duca degli Abruzzi 24, 10129 Torino, Italy\\
$^5$ School of Mineral Resources Engineering, Technical University of Crete
Chania, 73100, Greece\\
$^6$ Istituto dei Sistemi Complessi, Consiglio Nazionale delle Ricerche, c/o Dipartimento di Scienza Applicata e Tecnologia, Politecnico di Torino, Corso Duca degli Abruzzi 24, 10129 Torino, Italy\\
$^7$ Region of Electrical and Electronic Systems Engineering, Ibaraki University, 316-8511, Japan}

\date{\today}
\begin{abstract}
A great variety of complex physical, natural and artificial systems are governed by statistical distributions,  which often follow a standard exponential function in the bulk, while their tail obeys the Pareto power law. The recently introduced $\kappa$-statistics framework predicts distribution functions with this feature. A growing number of applications  in different fields of investigation are beginning to prove the relevance and effectiveness of $\kappa$-statistics
in fitting empirical data. In this paper, we use  $\kappa$-statistics to formulate a statistical approach for epidemiological analysis. We validate the theoretical results by fitting the derived $\kappa$-Weibull distributions with data from the plague pandemic of 1417 in Florence as well as data from the COVID-19 pandemic in China over the entire cycle that concludes in April 16, 2020. As further validation of the proposed approach we present a more systematic analysis of COVID-19 data from countries such as Germany, Italy, Spain and United Kingdom, obtaining very good agreement  between theoretical predictions and empirical observations. For these countries  we also study the entire first cycle of the pandemic which extends until the end of July 2020.
The fact that both the data of the Florence plague and those of the Covid-19 pandemic  are successfully described by the same theoretical model, even though the two events  are caused by different diseases and they are separated by more than 600 years, is evidence that the  $\kappa$-Weibull model has universal features.
\end{abstract}
\maketitle

\sect{Introduction}

Many phenomena in a large variety of disciplines present apparent regularities described by well-known statistical distributions \cite{martin2012}. Some examples include the populations of cities, the frequencies of words in a text, the energy distribution in solids, the behaviour of financial markets, the statistical distribution of the kinetic energy in a gas, etc.~\cite{Sornette06,West17}.
Moreover, statistical approaches provide  powerful tools for medical and epidemiological applications, since they allow predicting the behaviour of certain diseases~\cite{gc98,Daley99}. Thus, such approaches are very useful for informing health policy and decision making, particularly regarding control and  mitigation measures in response to the societal impacts of epidemics and pandemics.

While an epidemic is defined as ``the occurrence in a community or region of cases of an illness clearly in excess of normal expectancy'' \cite{madhavetal2017,portadict}, a pandemic is defined as ``an epidemic occurring over a very wide area, crossing international boundaries, and usually affecting a large number of people'' \cite{madhavetal2017,portadict}. Pandemics are large-scale outbreaks of infectious diseases which cause a growth in mortality over a wide geographic area.

In addition to the obvious health consequences, both epidemics and pandemics also cause economic stress and social hardship. It has been emphasized that the probability of occurrence of pandemics is increasing due to the high inter-connectedness of the modern world, which is facilitated by the ease of travel and a continuous increase of urbanisation \cite{madhavetal2017,jonesetal2008,morse1995}. Consequently, the international community is increasing its efforts toward the mitigation of the impacts of pandemics \cite{madhavetal2017}.
Some recent examples of pandemics are the 2003 SARS (Severe Acute Respiratory Syndrome), the 2014 West Africa Ebola epidemic, and the present COVID-19 (COronaVIrus Disease) pandemic caused by the SARS-CoV-2 virus.

In this paper we formulate a statistical thermodynamic approach to epidemiology, which demonstrates the utility of  $\kappa$-statistics for the analysis of epidemics and pandemics. First, we must consider certain facts which form  the biomedical base of these phenomena. Below we summarise the principle causes of pandemics \cite{madhavetal2017}:

\begin{itemize}
	\item	Pathogens such as influenza viruses that are capable of efficient transmission among humans, with a high potential to cause global and severe pandemics. They are characterized by a relatively long and asymptomatic infectious period, during which it is not possible to detect the infected persons and their movements.
	\item		Pathogens such as Nipah virus which are capable of generating a moderate global threat. They are transmitted efficiently as a result of mutations and adaptation.
	\item		Pathogens such as Ebola that are capable of causing local epidemics, with a non-negligible risk of evolving into a global pandemic.
\end{itemize}
Recently, pandemics have also found their origin in zoonotic transmissions of pathogens from animals to humans \cite{woolhouseetal}.

The risk of pandemic spread is conditioned by the following factors~\cite{sandsetal2016}:
\begin{itemize}
	\item	pathogen factors such as genetic adaptation and mode of transmission;
	\item	human factors such as population density, susceptibility to infection, travel, migration, poverty, malnutrition, and caloric deficits;
	\item	factors related to public policy such as public health surveillance and measurements.
\end{itemize}

In order to better predict and mitigate the societal impact, the ability to  quantify the morbidity and mortality associated with pandemics is very important. Consequently,  statistical approaches based on available data that  can lead to accurate models for predicting the behaviour of future pandemics are very useful. Historical data from past pandemics play a fundamental role, because they enable comparisons with theoretical models  and can also be used to assess the performance of model-based forecasts. Historical records are often sparse and incomplete. Nonetheless, in all fields of research, scientists and the engineers always search for ``optimal'' statistical distributions that can reliably predict the behaviour of different natural, engineered,  and social systems~\cite{
lucia2008,lucia2009,lucia2020
}.

Since the beginning of the Covid-19 epidemic, various works in the scientific literature have presented statistical analyses of the epidemiological data. Most of these works focus on the number of infections \cite{Tsallis2020,Vasconcelos,Zlatic,Piccolomini,Bastos,Perc,Pluchino} while the fatality curves  have only been analyzed in a few studies. The analysis in the paper~\cite{Vasconcelos} focuses on the cumulative data (cumulative distribution function) of Covid-19, using the Richardson growth model which depends on three free parameters. The data related to Covid-19 deaths per day (i.e., the empirical probability density function) are analyzed in the paper~\cite{Tsallis2020} by adopting a four-parameter model which is based on non-extensive statistics. However, to our knowledge the development of a statistical model that admits explicit expressions for both the mortality cumulative distribution and the respective probability density function remains an open problem.

A second question concerns the mathematical features of suitable statistical models and in particular the behavior of the tails of the distribution functions. All of the existing models depend on three to five free parameters. However, a good  fit of the data with a statistical model that depends on various free parameters is not sufficient proof of model validity. Two different model classes are mainly used to describe the Covid-19 data. The statistical distributions in the first class  include the Weibull, Richardson, and SIR (Susceptible-Infected-Recovered) model functions. These depend on three to five free parameters and exhibit exponential or stretched exponential tails. The second class  includes  statistical distributions with two, three, or four free parameters that feature power-law tails. Models in this class include the Pareto, Zipf, Burr, Student and Tsallis distributions.

The present work aims to address the two problems that are described above, i.e., \emph{analytical tractability} and \emph{tail behavior}. We propose a statistical model for the analysis of the fatality curves that originated in physics but has also been successfully used in other fields such as econophysics, finance, and seismology. This model depends only on  three free parameters but nonetheless provides excellent fits to the empirical curves obtained from the data.
In addition, the model admits simple, closed-form analytic  expressions for all the main statistical functions such as the probability density function, the cumulative distribution function, the survival function, the quantile function, the hazard function and the cumulative hazard function. Hence, this model overcomes the problem of analytical tractability and allows straightforward analysis of both daily and cumulative empirical data.

Regarding the second problem (tail behavior), the proposed $\kappa$-Weibull model   is closely related to both classes mentioned above, but it does not strictly belong in either class. Indeed, the model's distribution function  is a one-parameter continuous deformation of the Weibull distribution, which belongs in the first model class. However, the tail of the deformed distribution follows Pareto's power law as the distributions in the second class.  Two of the three model parameters,  namely $\alpha$ and $\beta$, correspond to the standard Weibull  parameters.  The third parameter, $\kappa$,  is related to deformation of  the tail of the distribution into a  power law; $\kappa$ is directly linked to the Pareto exponent $n$ of the survival function through the simple relationship $\kappa = \alpha /n$.

In order to test the validity of the $\kappa$-Weibull model we analyze various epidemics with different characteristics.
First, we analyze the mortality data from the Florence plague that occurred in 1417~\cite{kohn2008,cohonjr2004,delpanta}. These historical mortality data are of particular interest, because they were carefully recorded  and allow testing the model with an epidemic process that evolved over several months.
A second test of the $\kappa$-Weibull model involves the diffusion of  Covid-19 in China~\cite{link1,link2}. The mortality curve derived from the Chinese Covid-19 data becomes flat after April 15, 2020. On the other hand, on April 17, 2020 China reported 1290 additional deaths which occurred during the entire cycle of the epidemic. However, no details were given regarding the temporal distribution of the fatalities. Hence, these data cannot be incorporated in our analysis. After April 15, 2020 only two cases of death have been reported to date (both in May 2020). However, these concerned infected individuals who had returned to China from abroad.  For this reason, it is considered that the mortality cycle from Covid-19 in China concluded on April 16. The Chinese data analyzed herein are the
``partial data'' until April 16, 2020 which exclude the additional 1290 deaths that were \emph{a posteriori} reported on April 17.

To further test the $\kappa$-Weibull model, we present a  systematic analysis of COVID-19 data from other countries such as Germany, Italy, Spain and United Kingdom~\cite{link1,link2}. The analyzed data from these countries spans the time window from the beginning of the epidemic in February 2020  until the beginning of August 2020. This temporal range completely captures the entire cycle of the first wave of the epidemic.
Florence's data are completely different from those of Covid-19 in terms of the epidemiological cause, the duration, the fatality rate and the historical time. The fact that the $\kappa$-Weibull  model can successfully capture the behavior of data so different as those of Florence and Covid-19, as well as data from other fields of science (economics, finance, seismology) indicates that it possesses \emph{universality features} which make it suitable for a wide range of applications.

\sect{Methods}

The fundamental difficulty faced by mathematical approaches to epidemiology is that all forecasts  strongly depend on the model employed, the parameter estimates, as well as the choices of the initial conditions.
There are both deterministic and stochastic epidemiological models~\cite{Daley99,Chen05}.  The $\kappa$-deformed model that we propose herein has its roots in statistical mechanics and is therefore  stochastic by construction.

\subsect{The $\kappa$-deformed exponential function}
Let us consider the function $n(t)$ to represent the number density of deaths at any given time $t \in {\mathcal D}$, where ${\mathcal D}$ is the temporal domain of interest.  In most cases of interest ${\mathcal D} = [t_{0}, \infty )$, where $t_{0}\ge 0$.  The probability of death within a short time interval $\delta t$  around $t$ is given by $f(t) \delta t$, where $f(t)$ is the \emph{probability density function (pdf)}.   Then, the respective number density is given by $n(t)= N f(t)$, where $N$ is the total number of deaths.

In statistical mechanics a general rate equation for $f(t)$ is the first-order linear ordinary differential equation (ODE)
\begin{eqnarray}
\frac{d f(t)}{dt}=-r(t)\,f(t),
\label{1}
\end{eqnarray}
where the function $r(t)$ is the \emph{decay rate}.
The solution of the above ODE is the exponential
\begin{eqnarray}
f(t)=c \exp \left (-\int_{t_{0}}^{t} r(t')\,dt' \right )
\ ,\label{2}
\end{eqnarray}
with the standard normalisation condition which determines the constant $c$:
\begin{eqnarray}
\int_{t_{0}}^{t}  f(t')\,dt'=1 \ . \label{3}
\end{eqnarray}
The normalization also enforces a constraint on the number density, i.e.,  $N= \int_{t_0}^{\infty} n(t) dt.$

In the context of the exponential solution, the following three simple cases must be considered.
\begin{itemize}

	\item	The Exponential Model: this model is fundamental in every branch of science; indeed, it allows us to describe a great variety of phenomena, from elasticity to electricity, from nuclear decay to thermal transient response, to name a few. The exponential model  can be obtained for constant decay rate, i.e.,
	\begin{eqnarray}
	r(t)= \beta \ , \label{4}
	\end{eqnarray}
	which leads to the following exponential pdf:
	\begin{eqnarray}
	f(t)=\beta\,\exp(-\beta\,t) \ , \label{5}
	\end{eqnarray}
with ${\cal D}=(0,\,+\infty)$.

	\item Power-law Model (Pareto distribution): Zipf \cite{zipf1932,zipf1935,zipf1940,zipf1949}, in his studies on the size distribution of cities, incomes and word frequencies, pointed out the notion of regularity in the distribution of sizes. In a great variety of these cases the distributions follow a power law with an exponent close to $-1$,  also known as Zipf's Law \cite{wyllys1981,perline1996,okuyamaetal1999,li2002,newman2005,benguiguietal2011,piantadosi2014}. Such power law distributions have been considered with increasing interest in the description of regular distributions.  Distributions with exponents different from $-1$ are known as Pareto's law \cite{newman2005,kleiberetal2003,West17}.  In general, distributions with power-law tails are known as \emph{heavy-tailed, fat-tailed} or \emph{subexponential} distributions, in juxtaposition to distributions whose tails decay exponentially~\cite{Sornette06}.

	Perline introduced the following classification of power laws \cite{perline2005}:

		- A power law is \emph{strong} if the distribution is a power law over the entire domain of definition;\\
		- A power law is \emph{weak} if only part of the distribution is fitted by a power law;\\
		- A power law is \emph{false} if only a highly truncated part of the distribution is approximated by a power law (in the scientific literature there are many examples  of false power laws~\cite{Clauset09}).\\

	Pareto obtained his \textit{Type I model}, named \textit{the Pareto law} by fitting the  available data in his time on increasing social inequalities.  He concluded that only economic growth can increase the income of the poor and decrease inequality \cite{pareto1896}. Today, we know that his conclusions were partially right and partially wrong, because economic growth and equity are strictly related to the social relations of production, to the technological level and institutional structures of the economy, and to the composition, accumulation and distribution of human capital, as well as the quality and accessibility of the educational and financial structures in place \cite{dagum}. Moreover, since Pareto disseminated his approach in 1897 \cite{pareto1897}, the application of heavy-tailed distributions in economics has been developed \cite{toda2012}.
	
The Pareto pdf $f(t)$ is obtained from Eq. (\ref{1}) by inserting the following time-dependent decay rate function			

\begin{eqnarray}
	r(t)= \frac{p}{t}, \; p>1 \ , \label{6}
	\end{eqnarray}
	which leads to a power-law solution for the pdf, i.e.,
	\begin{eqnarray}
	f(t)=\frac{p-1}{t_0}\left ( \frac{t_0}{t}\right )^{p}, \; p>1 \ , \label{7}
	\end{eqnarray}
in the domain ${\cal D}=(t_0,\,+\infty)$, with $t_0>0$:

\item $\kappa$-Exponential Model: This model, which is based on a fundamental approach derived from relativity 
    \cite{Kaniadakis2002},  has proved useful in many applications. Experimental evidence suggests that probability density functions should resemble the exponential function for $t \rightarrow 0$. However, for $t \rightarrow 0$ the Pareto pdf diverges. On the other hand, for high values of $t$ many experimental results show a Pareto-like pdf with power law tails instead of exponential decay. Consequently, for $t \rightarrow 0$ it follows that $r(t) \sim \beta$ while for $t \rightarrow + \infty$ it follows $r(t) \sim p/t$.
So, the actual decay rate function $r(t)$ should smoothly interpolate between these two regimes; a good proposal for $r(t)$ has been introduced in the context of special relativity, where the function $r(t)$ is given in terms of the Lorentz factor. We recall the expression of the Lorentz factor $\gamma_{\kappa}(q)=\sqrt{1+\kappa^2\,q^2}$; this expression involves the dimensionless momentum $q$ where the parameter $\kappa$ is the reciprocal of the dimensionless light speed $c$, i.e. $\kappa \propto 1/c$. After posing $r(t)=\beta/\gamma_{\kappa} (\beta\,t)$ or more explicitly
	\begin{eqnarray}
	r(t)=\frac{\beta}{\sqrt{1+\kappa^2\,\beta^2\,t^2}} \ ,\label{8}
	\end{eqnarray}
it follows that for $t \rightarrow 0$ the decay rate $r(t)$ approaches the exponential regime, i.e. $r(t)\sim\beta$. On the other hand, for $t\rightarrow +\infty$ it follows that $r(t)$ approaches the decay rate of the Pareto model, i.e.  $r(t) \sim 1/\kappa\,t$.
	
	The solution of the rate equation in this case yields the following pdf
	\begin{eqnarray}
	f(t)=(1-\kappa^2)\,\beta\,\exp_{\kappa}(-\beta\,t) \ ,\label{9}
	\end{eqnarray}
	where the $\kappa$-deformed exponential function is given by
	\begin{eqnarray}
	\exp_{\kappa}(t)=\left(\sqrt{1+\kappa^2\,t^2}+\kappa\,t\right)^{1/\kappa} \ ,\label{10}
	\end{eqnarray}
	with $0<\kappa<1$. It is important to note that in the $\kappa \rightarrow 0$ limit and in the $t\rightarrow 0$ limit the function $\exp_{\kappa}(t)$ approaches the ordinary exponential $\exp(t)$, i.e.
	\begin{eqnarray}
	&&\exp_{\kappa}(t)
	{\atop\stackrel{\textstyle\sim}{\scriptstyle {\kappa}\rightarrow{0}}}\exp(t) \ ,\label{11} \\
	&&\exp_{\kappa}(t)
	{\atop\stackrel{\textstyle\sim}{\scriptstyle t \rightarrow{0}}}\,\exp(t) . \label{12}
	\end{eqnarray}
	On the other hand the function $\exp_{\kappa}(-t)$ for $t \rightarrow +\infty$ presents a power-law tail, i.e.
	\begin{eqnarray}
	\exp_{\kappa}(-t)
	{\atop\stackrel{\textstyle\sim}{\scriptstyle t\rightarrow
			{+\infty}}}\,(2 \kappa t)^{-1/ \kappa}.  \label{13}
	\end{eqnarray}
Furthermore, the $\kappa$-exponential satisfies the following identity
\begin{eqnarray}
  \exp_\kappa(t)\,\exp_\kappa(-t)=1 \ , \label{14}
\end{eqnarray} in analogy with the standard, non-deformed, exponential.
\end{itemize}

The $\kappa$-exponential represents a very powerful tool which can be used to formulate a generalized statistical theory capable of treating systems described by distribution functions that exhibit power-law
tails \cite{
Kaniadakis2002
,Kaniadakis2009
,PRE2017
}. The mechanism generating the $\kappa$-exponential function is based on first principles from special relativity, and therefore the new function appears  very promising for physical applications. Generalized statistical mechanics, based on the $\kappa$-exponential, preserves the main features of ordinary Boltzmann-Gibbs statistical mechanics which is based on the ordinary exponential through the Boltzmann factor. For this reason, it has attracted the interest of many researchers over the last two decades who have studied its foundations and mathematical aspects \cite{Silva06A,Naudts1,Topsoe,Tempesta2011,Scarfone2013,SouzaPLA2014,Scarfone1,Scarfone2,Gomez2020,Silva2020}, the underlying thermodynamics \cite{Wada1
,Bento3lawThermod
}, and specific applications of the theory to various fields. A non-exhaustive list of applications includes those in quantum statistics \cite{Santos2011a,Planck,Lourek2,Soares2019}, in quantum entanglement  \cite{Ourabah,OrabahPhyscripta}, in plasma physics \cite{Lourek,Gougam,Chen,Landau2017,Qualitative2017,Lourek2019,Khalid2020}, in nuclear fission \cite{NuclearEnergy2017,NuclearEnergy2018,Abreu2020,Shen2019},  in astrophysics \cite{Carvalho,Carvalho2,Carvalho2010,Cure}, in quantum gravity \cite{AbreuEPL,AbreuIJMPA,ChinesePL,AbreuEPL2018A,AbreuEPL2018B,Immirzi},  in geomechanics \cite{Oreste2017,Oreste2019}, in genomics \cite{SouzaEPL2014,Costa2019}, in complex networks \cite{Macedo,Stella,lei2020}, in economy \cite{Clementi2011,Bertotti,Modanese,Bertotti2017,Vallejos2019}, in finance \cite{Trivellato2012,Trivellato2013,Tapiero,Trivellato2017}, as well as in reliability analysis and seismology \cite{seismos,weak15,daSilva2020}.


\subsect{The $\kappa$-deformed statistical model}

Given a pdf $f(t)$ which represents the death rate, the \emph{cumulative distribution function (cdf)} $F(t):  {\mathcal D} \to [0, 1]$ represents the probability of death between the initial time and the current time $t$. $F(t)$, which is also known as the \emph{lifetime distribution}, is given by means of the following integral $F(t) = \int_{t_{0}}^{t} f(t') dt'.$

Conversely, the pdf $f(t)$ is given by the derivative of $F(t)$. In the following, we will assume without loss of generality that $t_{0}=0$.
The complement of $F(t)$, i.e., $S(t) = 1 - F(t)$ is known as the \emph{survival function}, and it represents the probability of survival at time $t$.

In most models of population dynamics the following time-dependent monomial is introduced
\begin{eqnarray}
T(t)=\beta\,t^{\alpha}\ .\label{15}
\end{eqnarray}
The expression (\ref{15}) for  $T$ contains the real-valued parameters $\alpha >0$ and $\beta >0$.
We can think of $T(t)$ as the time measured by a \emph{nonlinear clock}~\cite{Cushman09}. $T(t)$
is regularly used in the definition of  the survival function $S(t)$ which becomes an implicit function of time based on the dependence of $S$ on  $T=T(t)$, i.e. $S=S(T)$.

The survival function of the \emph{Weibull model} is then defined according to
\begin{eqnarray}
S=\exp(-T) \ ,\label{16}
\end{eqnarray}
and it represents a stretched exponential function in time.

Popular models for empirical data that exhibit power-law tails include the Log-Logistic model \cite{Collett2003}, with survival function given by
\begin{eqnarray}
S=\frac{1}{1+T} \ ,\label{17}
\end{eqnarray}
the Burr type XII or Singh-Maddala model \cite{Singh1976}, with survival function given by
\begin{eqnarray}
S=\frac{1}{(1+T)^p}, \quad p>0 \ ,\label{18}
\end{eqnarray}
and the Dagum model \cite{Dagum1977}, with survival function given by
\begin{eqnarray}
S=1-\frac{T^p}{(1+T)^p}, \quad p>0 \ .\label{19}
\end{eqnarray}

Various other  postulates can be used for the analytical expression of the survival function. However, for physical applications it is extremely important to identify physical mechanisms or first principles which lead to such expressions for the survival function.

The $\kappa$-deformed statistical model can be viewed as a one-parameter generalization of the Weibull model, obtained by replacing  the ordinary exponential $\exp(t)$ in the  definition (\ref{16}) of the survival function by the $\kappa$-deformed exponential $\exp_\kappa(t)$.   Then, using the Weibull dependence for $T$ given by Eq. (\ref{15}), we obtain the following expression for the $\kappa$-deformed survival function $S_\kappa=S_\kappa(t)$
\begin{eqnarray}
S_{\kappa}=\exp_\kappa(-T) \ ,\label{20}
\end{eqnarray}
or more explicitly
\begin{eqnarray}
S_\kappa(t)=\exp_\kappa(-\beta\,t^\alpha), \; \alpha>0, \beta>0 \ .\label{21}
\end{eqnarray}
The $\kappa$-deformed survival function $S_\kappa(t)$  reduces to the ordinary survival function of the Weibull model in the $\kappa \rightarrow 0$ limit.
The rate equation obeyed by $S_{\kappa}$ is expressed in terms of $T$ as a linear first-order ODE
\begin{eqnarray}
\frac{d S_\kappa}{d T}=-\frac{1}{\sqrt{1+\kappa^2\,T^2}}\, S_\kappa \ ,\label{22}
\end{eqnarray}
with initial condition $S_{\kappa}(0)=1$.
The ODE (\ref{22}) represents an interpolation between the rate equation of the exponential model and that of the Pareto model; moreover, the functional form in Eq. (\ref{22}) is dictated by the first principles of special relativity.

The most important feature of the function $S_\kappa(t)$ is that it continuously interpolates  between a power-law tail for large $t\gg1$, i.e.,
\begin{equation}
S_\kappa(t) \sim (2\,\kappa\,\beta)^{-1/\kappa}\,\,t^{-\alpha/\kappa} \ , \label{23}
\end{equation}
and exponential dependence
\begin{equation}
S_\kappa(t)\sim\exp(-\beta\,t^\alpha) \ ,
\end{equation}
for $t\ll1$.

The lifetime distribution function $F_\kappa=F_\kappa(t)$ is given by the expression
\begin{eqnarray}
F_\kappa(t)=1-\exp_\kappa(-\beta\,t^{\alpha}) \ ,\label{25}
\end{eqnarray}
while the pdf $f_\kappa=f_\kappa(t)$, defined  $f_\kappa=d F_\kappa/dt$, is given by
\begin{eqnarray}
f_\kappa(t)=\frac{\alpha\,\beta\,t^{\alpha-1}}{\sqrt{1+\kappa^2\,\beta^2\,t^{2\alpha}}}\, \exp_\kappa(-\beta\,t^\alpha) \ . \label{26}
\end{eqnarray}
The rate equation for the survival function assumes the form of the first-order linear ODE
\begin{eqnarray}
\frac{d S_\kappa(t)}{d t}= -\lambda_\kappa(t)\,S_\kappa(t) \ , \label{27}
\end{eqnarray}
where $S_\kappa(0)=1$, and $\lambda_\kappa=\lambda_\kappa(t)$ is the \emph{hazard function (hazard rate)} defined through
\begin{eqnarray}
f_\kappa(t)= \lambda_\kappa(t)\,S_\kappa(t). \label{28}
\end{eqnarray}
Hence, the hazard function assumes the expression
\begin{eqnarray}
\lambda_\kappa(t)=\frac{\alpha\,\beta\,t^{\alpha-1}}{\sqrt{1+\kappa^2\,\beta^2\,t^{2\alpha}}} \ . \label{29}
\end{eqnarray}
The cumulative hazard function $\Lambda_\kappa=\Lambda_\kappa(t)$ is defined by means of the integral
\begin{eqnarray}
\Lambda_\kappa(t)=\int_0^t \lambda_\kappa(u)\,du \ ,
 \label{30}
\end{eqnarray}
and is linked with  $S_\kappa$ through
\begin{eqnarray}
\Lambda_\kappa(t)=-\ln S_\kappa(t) \ .\label{31}
\end{eqnarray}
After taking into account that the $\kappa$-exponential function (\ref{10}) can also be expressed in the form
\begin{eqnarray}
\exp_\kappa(x)=\exp\left(\frac{1}{\kappa} ({\rm arcsinh } (\kappa\,x)\right) \ ,\label{32}
\end{eqnarray}
the cumulative hazard function $\Lambda_\kappa=\Lambda_\kappa(t)$ assumes the following explicit expression
\begin{eqnarray}
\Lambda_\kappa(t)=\frac{1}{\kappa}\,{\rm arcsinh}\, (\kappa\,\beta\,t^\alpha) \ ,\label{33}
\end{eqnarray}
and in the $\kappa \rightarrow 0$ limits reduces to the standard Weibull cumulative hazard function $\Lambda(t)=\beta\,t^\alpha$. It is important to note that in the standard Weibull model the cumulative hazard function coincides with the function $T(t)$.

Finally, the \emph{quantile function} $Q_\kappa$ is defined as the inverse of the survival function follows  $S_\kappa=S_\kappa(t)$. By expressing the quantile function  in the form $t=Q_\kappa(S_\kappa)$, one easily obtains
\begin{eqnarray}
Q_\kappa(u)=\left(-\frac{1}{\beta}\,\ln_\kappa(u)\right)^{1/\alpha} \ ,\label{34}
\end{eqnarray}
where the $\kappa$-logarithm $\ln_\kappa(u)$ is the inverse function of $\exp_\kappa(u)$, i.e. $\ln_\kappa(\exp_\kappa(u))=\exp_\kappa(\ln_\kappa(u))=u$ and is given by
\begin{eqnarray}
\ln_\kappa(u)=\frac{u^{\kappa}-u^{-\kappa}}{2\,\kappa} \ .\label{35}
\end{eqnarray}
After observing that the function $\ln_\kappa(u)$ approaches the function $\ln(u)$ in the $\kappa \rightarrow 0$ limit, it follows that in the same limit $Q_\kappa(u)$ reduces to the quantile function of the standard Weibull model.

Regarding the meaning of the parameters $\alpha$, $\beta$ and $\kappa$, we observe that the first two are the same as in the Weibull model and therefore have the same meaning. In particular, $\alpha $ is the \emph{shape parameter} while $\beta $ is linked to the \emph{scale parameter} $\tau $ through the relationship $\beta = \tau^{-1/\alpha} $. The deformation parameter $\kappa$, on the other hand, is linked to the Pareto exponent $n$ of the survival function through the simple relationship $n=\alpha/\kappa$ and to the Pareto exponent $p$ of the pdf via $p=1+\alpha/\kappa$; the former follows from the asymptotic dependence of the survival function given by Eq. (\ref{23}), while the latter from the fact that $f(t) = - d S(t)/dt$.

The mode of the probability density function, i.e. the time $t_M$ where the pdf attains its maximum value can be easily obtained analytically as a function of the free parameters $\alpha$, $\beta$ and $\kappa$.
More explicitly, after setting the first time derivative of the probability density function equal to zero  we obtain the following biquadratic equation
\begin{eqnarray}
  T_M^4\,\kappa^4-\left[\alpha^2\,T_M^4+2\,(\alpha-1)\,T_M^2\right]\,\kappa^2+(\alpha-1)^2-\alpha^2\,T_M^2=0 \ ,
\end{eqnarray}
with $T_M=\beta\,{t_M}^\alpha$.\\
This equation can be solved to obtain $T_M$ as a function of $\alpha$ and $\kappa$. It can also be solved  to determine $\kappa$ as a function of $\alpha$ and $T_M$, i.e.,
\begin{eqnarray}
\kappa={1\over T_M}\sqrt{\alpha-1+{1\over2}\,\alpha^2\,T_M^2-\alpha\,T_M\,\sqrt{\alpha+{1\over4}\,\alpha^2\,T_M^2}} \ .
\end{eqnarray}
This last expression is very useful in the analysis of empirical data.

Finally,  the  $\kappa$-deformed statistical model presented above, which represents a one-parameter continuous deformation of the Weibull model, has been successfully applied  in econophysics for the analysis of personal income distribution \cite{Clementi2011} and in seismology  \cite{seismos}.  We believe that the applicability of the model to such complex dynamical systems which are quite different from each other suggests that the model involves universal features.

\sect{Results}

The purpose of this section is the validation of the $\kappa$-Weibull statistical model which is described in detail in the ``Methods'' section. The validation  is performed by applying the model to various pandemic data. An intriguing first application regards the analysis of the number of deaths during the plague pandemic that ravaged the city of Florence in the XV century.

The Bubonic Plague or Black Death arrived in Europe in 1348 and in Italy in the spring of the same year~\cite{kohn2008,cohonjr2004}. It should be noted that the plague inspired Boccaccio to write his famous nouvelle  \textit{Decameron} only a couple of years after the end of the pandemic. Its consequence was the death of around 25-50\% of Europe’s population by 1351. The pandemic is believed to have started in China and came to Europe via the trading routes though Asia and  the Black Sea. 
In the year 1417, Florence had a population of  60,000 inhabitants and the Italian city's Grain Office maintained a series of Books of Dead which recorded the number of human deaths caused by the bubonic plague. Florentines contracted the disease through contact with  infected black rats and the fleas that they carried~\cite{cohonjr2004}.  From a thermodynamic point of view, deaths which are due to the spread of an epidemic can be considered as an irreversible process inside an open system.
Herein we show how the evolution of the plague in relation to the time and the number of deceased people can be explained using the  $\kappa$-deformed statistics.  In Table 1 the statistical data of the 1417 plague in Florence are summarised as a function of time \cite{delpanta}.

\begin{table}[ht]
	\centering
	\begin{tabular}{l c c}
		\textbf{Month} & \textbf{Dead people} & \textbf{Population of Florence} \\ \hline
		May	& 600	& 59400 \\
		June &	700 &	58700 \\
		July &	2700 &	56000 \\
		August	& 5000	& 51000 \\
		September &	2000	& 49000 \\
		October	& 600	& 48400  \\
		November	& 200	& 48200 \\
		December	& 100 &	48100 \\ \hline
	\end{tabular}
\caption{Temporal distribution of plague victims and population in Florence during the year 1417 \cite{delpanta}}
\end{table}

\begin{figure}[ht]
	\centering
	\includegraphics[width=18 cm]{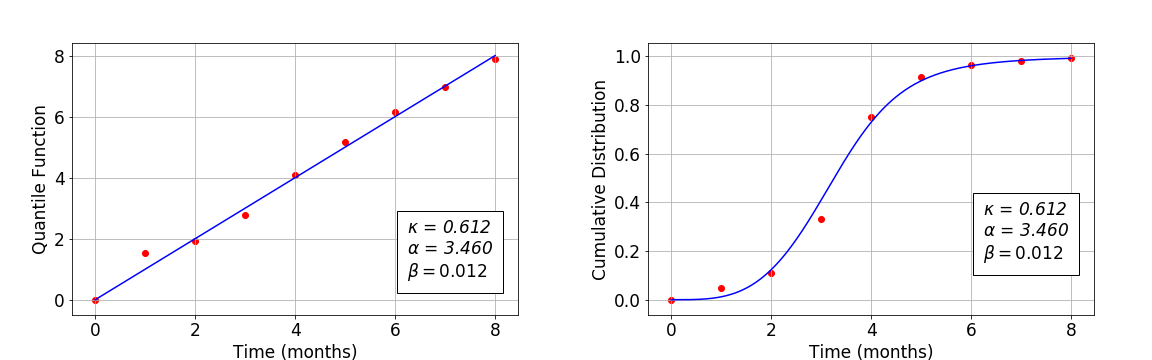}
	\caption{Theoretical (continuous curve) and empirical (dots) plots of the quantile function (left) and the cumulative distribution function (right) versus time for the 1417 Florence plague epidemic. The theoretical curves are based on Eq. (\ref{34}) and Eq. (\ref{25}), respectively.}
\end{figure}

We consider the \emph{death occurrence probability}  as the ratio between the monthly number of deaths (given by the second column of the Table 1) and the cumulative number of deaths. We assume that the latter is equal to the total number of deaths that were recorded between May and December increased by an additional 100 deaths; we hypothesize that the latter occurred in the months following December, raising the total number of deaths to $N=12000$.

Figure 1 shows the quantile function $Q_\kappa$ (left) and the cumulative distribution function $F_\kappa$ (right), versus the time $t$ for the Florence epidemic. The theoretical quantile function
of the $\kappa$-deformed model, defined in Eq. (\ref{34}), is just the bisectrix of the plane represented by the continuous straight line. The continuous sigmoid curve represents the theoretical cumulative distribution function as given by Eq. (\ref{25}). The empirical data related to the two functions (marked on the plots by dots) are deduced from the data in Table 1. The optimal fit parameters have the following values: $\kappa= 0.612$, $\alpha = 3.460$ and $\beta = 0.012$. It is clear that the model describes the empirical data very well. Deviations of the empirical quantiles from the theoretical curves are likely due to detection errors since as it is evidenced in Table 1, the number of monthly deaths is approximated with an error margin of the order of 100 units. This statistical error is responsible for the dispersion of empirical quantiles around the bisectrix. Hence, the optimal parameter estimate $\kappa = 0.612$ clearly shows the difference (supported by the data) between the optimal $\kappa$-deformed model and the standard Weibull model corresponding to value $\kappa=0$.

As second application we analyze the data of Covid-19 mortality in China during the winter and spring of 2020~\cite{link1,link2}. The Chinese data set on SARS-CoV-2 are important, as they represent the first statistical record for the entire evolutionary cycle of the COVID-19 pandemic in the world. It is important to note that the first and only cycle of Covid-19 ended in China with a total of 3,342 deaths around April 16, 2020. After April 17, 2020 while the Covid-19 cycle in China had already ended, the authorities reported 1290 additional deaths that occurred outside the hospitals during the entire cycle without disclosing their temporal distribution. For this reason the present analysis is applied to the 3,342 deaths which were reported by April 16, 2020.

\begin{figure}[ht]
	\centering
	\includegraphics[width=18 cm]{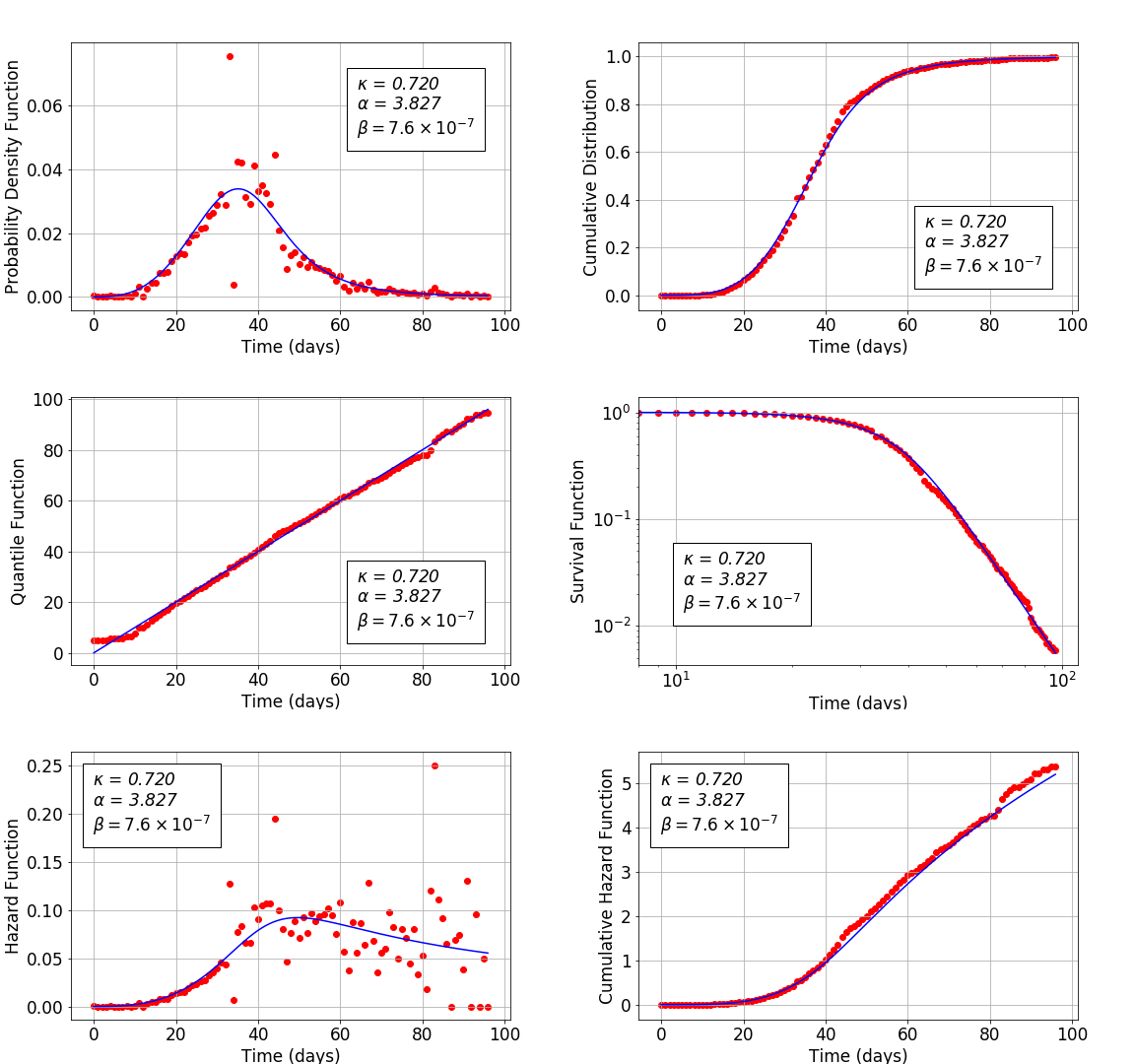}
	\caption{Theoretical (continuous curve) and empirical (dots) plots of the probability density function (top left), cumulative distribution function (top right), quantile function (middle left), survival function (middle right), hazard function (bottom left) and cumulative hazard function (bottom right) versus time for the Covid-19 mortality data related to China 2020 epidemic. The theoretical curves are based on Eq. (\ref{26}), Eq. (\ref{25}), Eq. (\ref{34}), Eq. (\ref{21}), Eq. (\ref{29}) and Eq. (\ref{33}) respectively.}
\end{figure}

Figure 2 reports the probability density function $f_\kappa$ (Eq. (\ref{26})), the cumulative distribution function $F_\kappa$ (Eq. (\ref{25})), the quantile function $Q_\kappa$ (Eq. (\ref{34})), the survival function $S_\kappa$ (Eq. (\ref{21})), the hazard function $\lambda_\kappa$ (Eq. (\ref{29})) and the cumulative hazard function  $\Lambda_\kappa$ (Eq. (\ref{33})) versus the time $t$ (measured in days) for the COVID-19 pandemic in China. The best fit parameters have the values: $\kappa= 0.720$, $\alpha= 3.827$ and $\beta= 7.6\cdot10^{-7}$. The plots in this figure confirm the goodness of fit of the $\kappa$-deformed model. In particular, there is no systematic deviation of the empirical quantiles from the theoretical linear trend while the statistical nature of data leads to slight dispersion around the bisectrix. Furthermore the survival function $S_\kappa$ versus time$t$ is shown on a log-log plot in order to better explore  differences between the empirical data and the theoretical predictions, especially in the tail of the distribution. The almost linear decay of the tail in this log-log plot is remarkable, showing clear signs that the Chinese Covid-19 data follows the Pareto power law in the distribution tail. The dispersion around the theoretical curves of the empirical data for both  $f_\kappa$ and $\lambda_\kappa$ reflects statistical fluctuations that are independent on the adopted theoretical model. These fluctuations are averaged out and become less evident in the cumulative functions $F_\kappa$, $S_\kappa$, $\Lambda_\kappa$ and $Q_\kappa$.

In the following we focus on the pandemic in Europe; in particular, we analyze the mortality Covid-19 data of Germany, Italy, Spain and United Kingdom, during the spring of 2020~\cite{link1,link2}. In these European countries the virus began to spread later than in China while the mortality exceeded the Chinese rate by more than 10 times.

\begin{figure}[ht]
	\centering
	\includegraphics[width=18 cm]{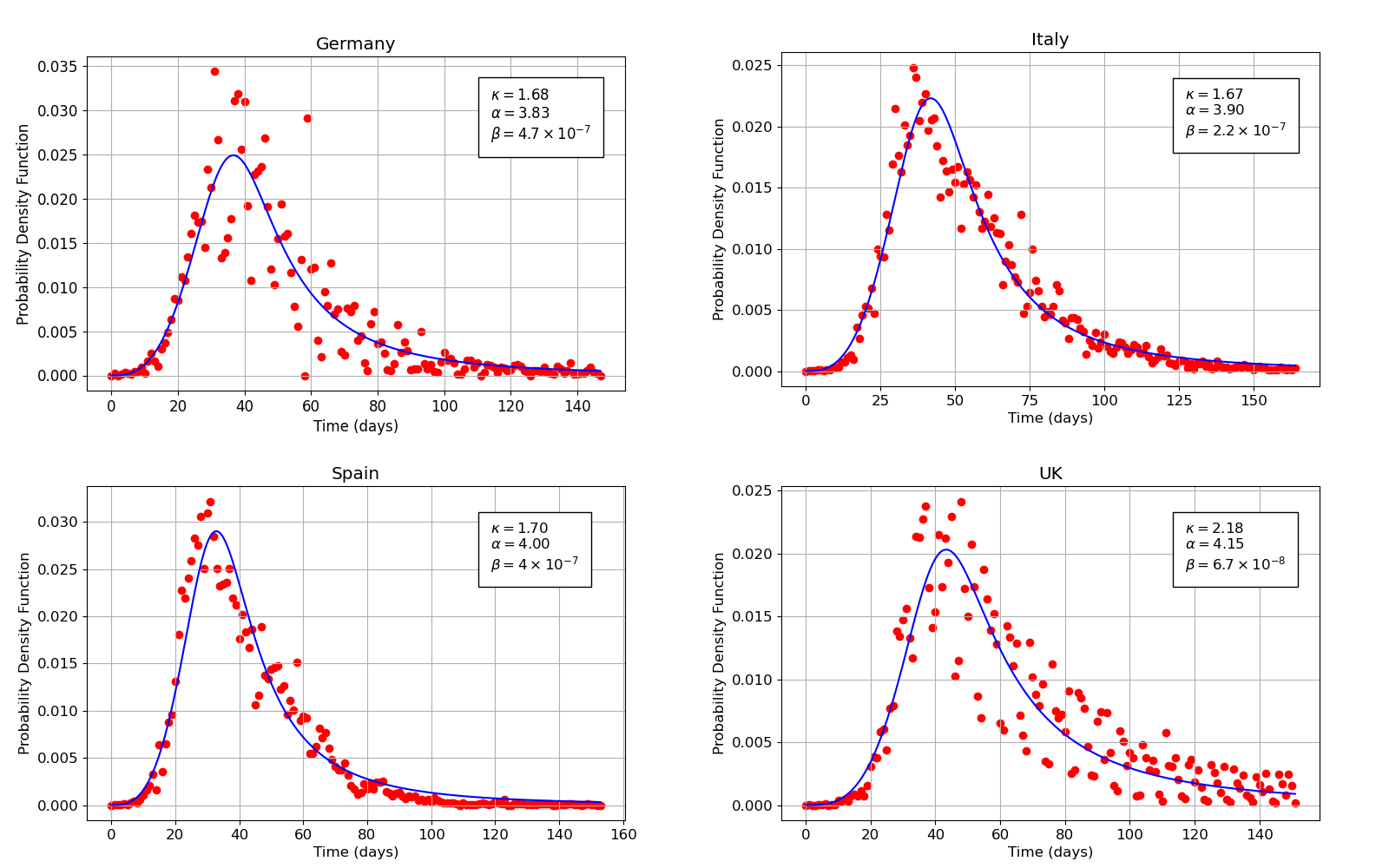}
	\caption{Theoretical (continuous curve) and empirical (dots) plots of the probability density function versus time for the Covid-19 mortality data related to Germany (top left), Italy (top right), Spain (bottom left) and United Kingdom (bottom right) 2020 pandemic. The theoretical curves are based on Eq. (\ref{26}).}
\end{figure}

Figure 3 shows the probability density function $f_\kappa(t)$ versus time related to Covid-19 mortality daily data for Germany, Italy, Spain and United Kingdom. The theoretical curve is given by  Eq. (\ref{26}). The parameter $\alpha$ for the four countries assumes the values $3.83$, $3.90$, $4.00$ and $4.15$ respectively. The parameter $\kappa$ for the four countries assumes the values $1.68$, $1.67$, $1.70$ and $2.185$ respectively. Finally the parameter $\beta$ for the four countries assumes the values $4.7\cdot10^{-7}$, $2.2\cdot10^{-7}$, $4.00\cdot10^{-7}$ and $6.7\cdot10^{-8}$ respectively. The comparison of the above optimal parameters for the four countries is performed below.

 \begin{figure}[ht]
	\centering
	\includegraphics[width=18 cm]{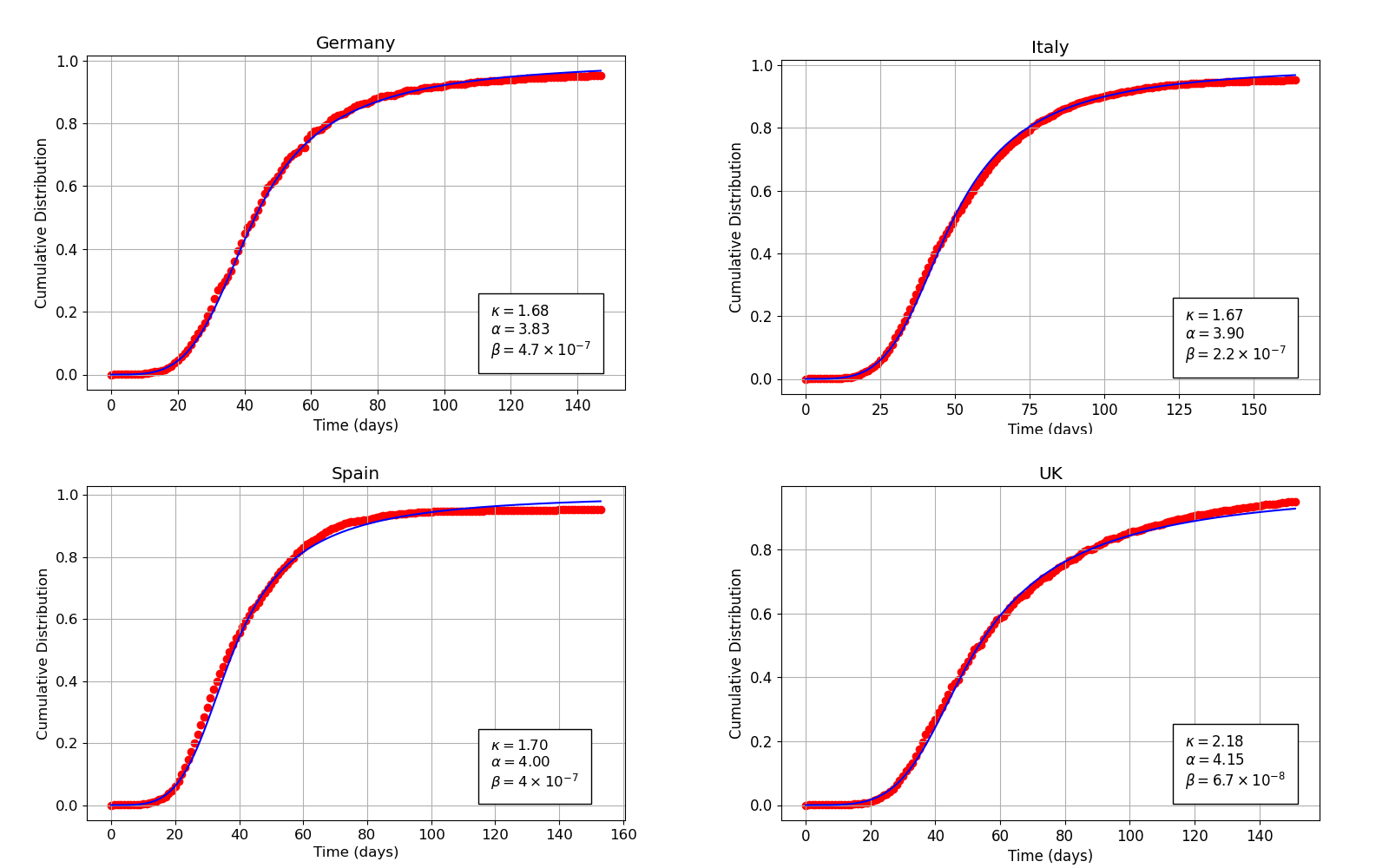}
	\caption{Theoretical (continuous curve) and empirical (dots) plots of the cumulative distribution function versus time for the Covid-19 mortality data related to Germany (top left), Italy (top right), Spain (bottom left) and United Kingdom (bottom right) 2020 pandemic. The theoretical curves are based on Eq. (\ref{25}).}
\end{figure}

Figure 4 shows the cumulative distribution function $F_{\kappa}(t)$ versus time related to Covid-19 mortality data for Germany, Italy, Spain and United Kingdom. The theoretical curve is given by  Eq. (\ref{25}). The optimal values of the parameter $\alpha$, $\kappa$ and $\beta$ for the four countries are the same as in Figure 3. It is worth noting that the statistical fluctuations of the daily data which are evident in Figure 3 are absorbed in the representation of the cumulative data shown in Figure 4.

\begin{figure}[ht]
	\centering
	\includegraphics[width=18 cm]{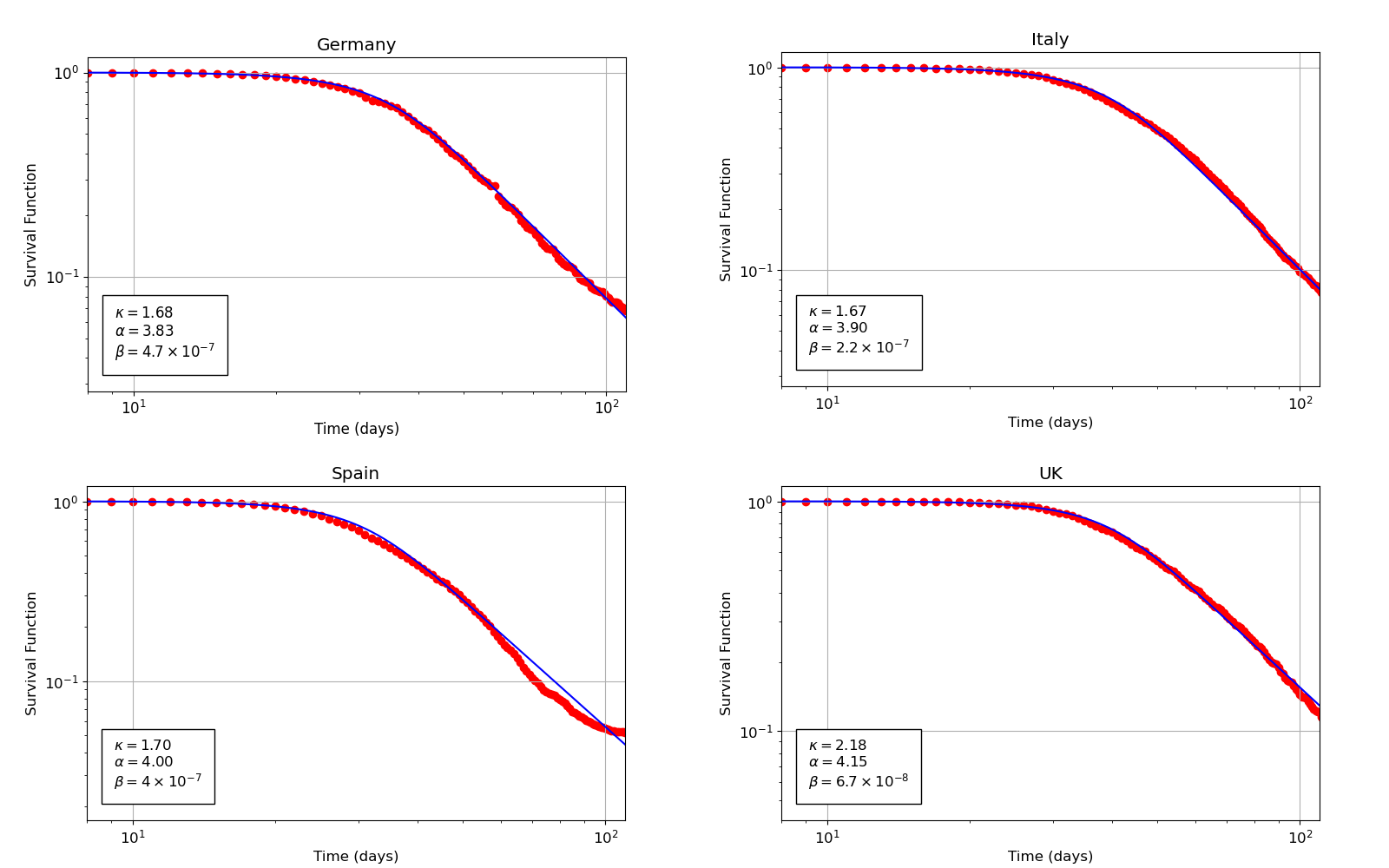}
	\caption{Theoretical (continuous curve) and empirical (dots) plots of the survival function versus time for the Covid-19 mortality data related to Germany (top left), Italy (top right), Spain (bottom left) and United Kingdom (bottom right) 2020 pandemic. The theoretical curves are based on Eq. (\ref{21}).}
\end{figure}

Figure 5 displays the survival function $S_\kappa(t)$ versus time on a log-log plot related to Covid-19 mortality data for Germany, Italy, Spain and United Kingdom. The continuous curve is the theoretical function defined in Eq. (\ref{21}). The optimal values of the parameter $\alpha$, $\kappa$ and $\beta$ for the four countries are the same as in Figure 3. In this log-log plot, a remarkable agreement between the theoretical predictions and the empirical data clearly emerges. Furthermore, the almost linear decay of the tail of the survival function shown in this plot indicates a Pareto power-law tail for all the  countries studied here, in agreement with the $\kappa$-deformed model.

For the purpose of comparing the Covid-19 mortality data from China, Germany, Italy, Spain and United Kingdom we focus on the probability density function. First, we discuss  the growth in mortality (i.e.,  the left tail) which is due to the initial phase of the virus spread. This is followed by a discussion of the final phase of the studied cycle, during which the decrease in the virus spread is described by the right tail of the curve.

The asymptotic behaviour of  $f_\kappa(t) $ for $t\rightarrow 0$ (left tail) is given by $f_\kappa(t) \propto t^\alpha$ where $\tau=\beta^{-1/ \alpha}$ is the \emph{Weibull scale parameter}. The shape parameter for China, Germany, Italy, Spain and United Kingdom has the values $3.83$, $3.83$, $3.90$, $4.00$ and $4.15$ respectively. These estimates of $\alpha$ are very close in value, with an average value equal to $3.94 \pm 0.2$.  It can therefore be concluded that for the five countries analyzed, the initial growing phase of the Covid-19 mortality presents quite similar dynamic behavior.

The asymptotic behaviour of $f_\kappa(t)$ for $t\rightarrow \infty$ (right tail) is given by $f_\kappa(t) \propto t^{-p}$ where $p=1 + \alpha / \kappa$ is the pdf's Pareto parameter (tail exponent). This parameter which governs the behavior of the right tail of the distribution takes the values $6.31$, $3.28$, $3.33$, $3.35$ and $2.83$  for China, Germany, Italy, Spain and United Kingdom respectively. It can be observed that the dynamics of the final phase of the first cycle of the epidemic is essentially the same for the three continental European countries in the study.  For these countries the Pareto parameter is approximately equal to $3.32\pm0.04$. For the United Kingdom, on the other hand, the Pareto parameter is lower (equal to $2.83$)  indicating that the tail is thicker, i.e., that mortality is more persistent over time. On the contrary,  the Paretto index is higher (equal to $6.31$) for the Chinese data which indicates a faster decay of the mortality curve than the respective European curves. The differences in the Pareto parameters between the different countries are likely to reflect differences in the response strategies, in the compliance of the citizens to restrictive measures that can curb the spread of the virus, as well as in the readiness and quality of the health care system including emergency treatment units.

\sect{Discussion}

Both the plague data from the pandemic of 1417 in Florence as well as the Covid-19 data of the 2020 pandemic from China, Germany, Italy, Spain and United Kingdom  have been analyzed by means of the proposed $\kappa$-Weibull model, to obtain information about the spreading dynamics of these deadly disease outbreaks.

It is worth recalling that the Florence plague data are approximate and relatively sparse, and thus they cannot be used to reliably test a statistical model. However, the Florence data are of great historical importance as they represent the first quantitative and regularly collected record of a pandemic. The first two figures in this paper undoubtedly show that at least the cumulative Florence data are compatible with the proposed  $\kappa$-deformed statistical model.

On the other hand, the Chinese data set on SARS-CoV-2 are important, as they represent  the first statistical record for the entire evolutionary cycle of the COVID-19 pandemic in the world. The Chinese mortality data have been successfully used to perform a first validation test of the $\kappa$-deformed model.

It is important to note that the first and only (so far) cycle of Covid-19 ended in China with a total of 3,342 deaths. This is a relatively low mortality compared to  other countries where the number of deaths was decidedly higher and mortality figures rose to a few dozen times that of China. A second noteworthy fact is that on April 17, 2020 while the Covid-19 cycle in China had already ended, the Chinese authorities reported 1290 additional deaths that occurred outside hospitals during the entire cycle without providing any information regarding their temporal distribution.

It is therefore evident that the data pertaining to the 3,342 deaths caused by Covid-19 in China are partial. For this reason, the validation of the $\kappa$-deformed model with complete data from other countries affected by Covid-19 is extremely important.
Thus, we focused on Europe and analyzed the Covid-19 data of Italy and Spain, where the virus began to spread towards the beginning of March 2020, as well as the data from Germany and the United Kingdom, where the virus began to spread two weeks later. These data from the European  countries and China differ with respect to the total number of deaths, which in the case of Germany is about 3 times the number of Chinese deaths while for Italy, Spain and United Kingdom the mortality is about 10 times that of China. Very good agreement has been obtained by fitting the $\kappa$-Weibull model to the data from these European countries, thereby providing further and more reliable validation of the theoretical model.

The $\kappa$-deformed, three-parameter model admits simple analytical forms for all the main statistical functions (probability density function, cumulative distribution function, survival function, quantile function, hazard function, cumulative hazard function) and can therefore be easily applied to Covid-19 data. Furthermore this model preserves all the universality properties already present in the original Weibull model. The results obtained by applying the $\kappa$-deformed model to mortality data from Covid-19 pertaining to the first cycle of the pandemic in China, Germany, Italy, Spain and United Kingdom, suggest that the $\kappa$-deformed model could also be used to model infection data from the first cycle of the Covid-19 pandemic as well as for the analysis of the second pandemic cycle which has already started in August 2020.

\bibliography{article_revtex}



\sect{Author contributions statement}

G.K. developed the statistical model and U.L. conceived the idea to adopt the model in the analysis of pandemic data. All authors provided critical feedback and helped shape the research, analysis and manuscript.

\sect{Additional information}

\textbf{Competing interests}: The authors declare no competing interests.\\
The corresponding author is responsible for submitting a \href{http://www.nature.com/srep/policies/index.html#competing}{competing interests statement} on behalf of all authors of the paper. 

\end{document}